\documentclass[twocolumn,english,showpacs]{revtex4}
\usepackage{babel,amsmath,amssymb,dcolumn}
\usepackage[dvips]{graphics}

\def\imo{i}
\begin{document}

\title{Perturbations and quasi-normal modes of black holes in
Einstein-Aether theory}

\author{R.A. Konoplya}
\email{konoplya@fma.if.usp.br}
\author{A. Zhidenko}
\email{zhidenko@fma.if.usp.br}
\affiliation{Instituto de F\'{\i}sica, Universidade de S\~{a}o Paulo \\
C.P. 66318, 05315-970, S\~{a}o Paulo-SP, Brazil}


\begin{abstract}
We develop a new method for calculation of quasi-normal modes of
black holes, when the effective potential, which governs black
hole perturbations, is known only numerically in some region near
the black hole. This method can be applied to perturbations of a
wide class of numerical black hole solutions. We apply it to the
black holes in the Einstein-Aether theory,  a theory where general
relativity is coupled to a unit time-like vector field, in order
to observe local Lorentz symmetry violation. We found that in the
non-reduced Einstein-Aether theory, real oscillation frequency and
damping rate of quasi-normal modes are larger than those of
Schwarzschild black holes in the Einstein theory.
\end{abstract}

\maketitle




\section{Introduction}

General Relativity is based on local Lorentz invariance, yet recently
there appeared a lot of attempts to go beyond local  Lorentz symmetry
\cite{LV}. Aether can be considered as locally preferred state of rest
at each point of space-time due-to some unknown physics. Einstein-Aether
theory is general relativity coupled to a dynamical time-like vector
field $u^{a}$, which is called ``aether''. This theory is what comes
instead of usual General Relativity when local Lorentz symmetry is
broken. Namely, $u^{a}$ breaks local boost invariance, while
rotational symmetry in a preferred frame is preserved (see
\cite{AEreview} for a recent review). Different observable
consequences of  Einstein-Aether theory have been discussed
recently \cite{AEobserve}.

Quasi-normal modes have been studies recent years extensively,
because of their interpretation in Conformal Field Theory
\cite{CFT} and possible, but yet unclear significance in Loop
Quantum Gravity. At the same time, one of the most promising ways
to check the theory of gravity as a fully non-linear theory is to
observe the characteristic frequencies of black holes called {\it
quasi-normal modes} \cite{QNMs}. QN frequencies are expected to be
observed in the nearest future with the help of a new generation
of gravitational antennas. This suggests a unique opportunity to
test the spontaneous breaking of the local Lorentz invariance,
through observing deviations of quasi-normal modes of black holes
from their values in Einstein theory.

Therefore, we would like to know what comes instead of
the well-known quasi-normal spectrum of black holes in
Einstein-Aether theory. For this one must have a solution
describing black holes in the Einstein-Aether theory. Fortunately such a
solution has been recently obtained in \cite{Jacobson}, yet, only
numerically. That was one of the motivations for us for developing  of
the method for finding quasi-normal modes for potentials which are not
known in analytical form, but are given only numerically in some
region near a black hole (see Sec. III of this paper). In case of
asymptotically anti-de Sitter space-times, the method for finding
QN modes for metrics given by a set of differential equations was
proposed in \cite{Zhidenko:2005mv}.

We found here that quasi-normal modes for spherically symmetric
black holes in the non-reduced Einstein-Aether theory have larger real
oscillation frequency and damping rate than the corresponding QN modes
in the locally Lorentz invariant Einstein gravity. 

\section{Basic formulas}

The lagrangian of the full Einstein-Aether theory forms the most
general diffeomorphism invariant action of the space-time metric
$g_{ab}$ and the aether field $u^a$ involving no more than two
derivatives given by
\begin{equation}\label{eaft}
L=-R-K^{ab}_{~~~mn}\nabla_au^m\nabla_bu^n-\lambda(g_{ab}u^au^b-1),
\end{equation}
here $R$ is the Ricci scalar, $\lambda$ is a Lagrange multiplier which provides the unit time-like constraint,
$$K^{ab}_{~~~mn}=c_1g^{ab}g_{mn}+c_2\delta^a_m\delta^b_n+c_3\delta^a_n\delta^b_m+c_4u^au^bg_{mn},$$
where $c_i$ are dimensionless constants.

Spherical symmetry allows to fix $c_4=0$. In this letter, following
\cite{Jacobson}, we shall consider the so-called {\it non-reduced} Einstein-Aether
theory,
for which $c_3=0$, and we can use the field redefinition that
fixes the coefficient $c_2$ \cite{Jacobson}:
$$c_2=-\frac{c_1^3}{2-4c_1+3c_1^2},$$
so that $c_1$ is the free parameter.

The metric for a spherically symmetric static black hole in
Eddington-Finkelstein coordinates can be written in the form
\cite{Jacobson}:
\begin{equation}
d s^{2} = N(r) d v^{2} - 2 B(r) d v d r - r^{2} d \Omega^{2},
\end{equation}
where the functions $N(r)$ and $B(r)$  are given by numerical
integration near the black hole event horizon \cite{Jacobson}.
One can re-write this metric in a Schwarzschild like form:
\begin{equation}\label{metric}
d s^{2} = - N(r) d t^{2} + \frac{B^{2}(r)}{N(r)}d r^{2} + r^{2} d \Omega^{2}.
\end{equation}

Now, let us consider test scalar and electromagnetic fields in the
background of the black hole given by the metric (\ref{metric}).
Here we shall use general covariant generalizations of scalar and
electromagnetic wave equations, i.e. we neglect interaction of
these fields with aether. This can be well understood, since the
background of a large astrophysical black hole has much more
influence on propagation of fields during quasi-normal ringing
than aether, at least not very far from the black hole. Yet, such
phenomena as tales at asymptotically late times are probably
affected by asymptotic regions of space-times as well as by a
black hole itself.

Thus, the wave equations for test scalar $\Phi$ and
electromagnetic $A_{\mu}$ fields are:
\begin{equation}
(g^{\nu \mu}
\sqrt{-g} \Phi_{,\mu})_{,\nu} = 0,
\end{equation}
\begin{equation}
((A_{\sigma, \alpha} - A_{\alpha, \sigma}) g^{\alpha \mu} g^{\sigma \nu}
\sqrt{-g})_{, \nu} = 0.
\end{equation}
After making use of the metric coefficients (\ref{metric}), we found that
the perturbation equations can be reduced to the wave-like form for
the scalar and electromagnetic wave functions $\Psi_{s}$ and $\Psi_{el}$:
\begin{equation}\label{wavelike}
\frac{d^{2} \Psi_{i}}{d r_{*}^{2}} + (\omega^{2} - V_{i}(r))\Psi_{i} =
0,\qquad d r_{*}=\frac{B(r)}{N(r)}dr.
\end{equation}
The effective potentials take the form:
\begin{equation}\label{sp}
V_{s} = N(r)\frac{\ell (\ell + 1)}{r^{2}} + \frac{1}{r}
\frac{d} {d r_*}\frac{N(r)}{B(r)},
\end{equation}
\begin{equation}\label{emp}
V_{el} = N(r) \frac{ \ell (\ell + 1)}{r^{2}}.
\end{equation}

Below we shall also use the particular case of the above metric
when $B(r)=1$ and $N(r)=1-2M/r$, what corresponds to an ordinary
Schwarzschild black hole. Now we are in position to find
quasi-normal spectrum for the above wave equations. Yet, as the
function $B(r)$, and $N(r)$ are known only numerically, in the
next section we shall have to develope a new technique for finding
of quasi-normal modes in this case.

\section{A new method for calculation of quasi-normal modes for potentials unknown analytically}


\begin{table*}

\caption{Schwarzschild fundamental quasi-normal modes of electromagnetic
  perturbations calculated by different
  methods: by using exact potential, using interpolation of the
  potential near its peak by 20 and 50 points per 2M unit, using fit
  of the metric functions by 10 and 100 points found with
  1\% precision.\\}\label{edata}

\begin{tabular}{|r|c|c|c|c|c|}
\hline
s=1&Exact potential&20 points interpol.&50 points interpol.&Fit
(10 points, 1\%)&Fit (100
points, 1\%)\\
\hline
$\ell=1$,WKB3&0.491740-0.186212\imo&0.491669-0.186217\imo&0.491739-0.186216\imo&0.489680-0.184993\imo&0.491686-0.186181\imo\\
$\ell=1$,WKB6&0.496383-0.185274\imo&0.512474-0.174892\imo&0.496625-0.184945\imo&0.494266-0.184069\imo&0.496328-0.185244\imo\\
\hline
$\ell=2$,WKB3&0.914262-0.190130\imo&0.914225-0.190139\imo&0.914261-0.190132\imo&0.910125-0.188874\imo&0.914155-0.190099\imo\\
$\ell=2$,WKB6&0.915187-0.190022\imo&0.916133-0.189337\imo&0.915134-0.190010\imo&0.911038-0.188768\imo&0.915080-0.189991\imo\\
\hline
$\ell=3$,WKB3&1.313467-0.191262\imo&1.313441-0.191271\imo&1.313466-0.191262\imo&1.307435-0.189994\imo&1.313311-0.191230\imo\\
$\ell=3$,WKB6&1.313797-0.191234\imo&1.313923-0.191101\imo&1.313799-0.191231\imo&1.307761-0.189967\imo&1.313641-0.191202\imo\\
\hline
$\ell=4$,WKB3&1.706036-0.191730\imo&1.706016-0.191740\imo&1.706036-0.191730\imo&1.698158-0.190457\imo&1.705833-0.191698\imo\\
$\ell=4$,WKB6&1.706190-0.191720\imo&1.706205-0.191683\imo&1.706190-0.191720\imo&1.698310-0.190448\imo&1.705987-0.191688\imo\\
\hline
$\ell=5$,WKB3&2.095741-0.191968\imo&2.095725-0.191978\imo&2.095741-0.191968\imo&2.086038-0.190693\imo&2.095492-0.191936\imo\\
$\ell=5$,WKB6&2.095826-0.191963\imo&2.095818-0.191953\imo&2.095826-0.191963\imo&2.086121-0.190688\imo&2.095575-0.191931\imo\\
\hline
\end{tabular}

\caption{The same as in the table \ref{edata} for scalar field.}\label{sdata}

\begin{tabular}{|r|c|c|c|c|c|}
\hline
s=0&Exact&20 points int.&50 points int.&Fit (10 points,1\%)&Fit
(100
points 1\%)\\
\hline
$\ell=1$,WKB3&0.582228-0.196003\imo&0.582003-0.195801\imo&0.582219-0.195995\imo&0.579382-0.194684\imo&0.582155-0.195970\imo\\
$\ell=1$,WKB6&0.585819-0.195523\imo&0.589883-0.192359\imo&0.585940-0.195409\imo&0.582932-0.194211\imo&0.585745-0.195490\imo\\
\hline
$\ell=2$,WKB3&0.966422-0.193610\imo&0.966365-0.193592\imo&0.966420-0.193610\imo&0.961834-0.192317\imo&0.966304-0.193577\imo\\
$\ell=2$,WKB6&0.967284-0.193532\imo&0.968078-0.192999\imo&0.967284-0.193532\imo&0.962685-0.192240\imo&0.967166-0.193500\imo\\
\hline
$\ell=3$,WKB3&1.350412-0.193024\imo&1.350384-0.193029\imo&1.350411-0.193025\imo&1.344061-0.191738\imo&1.350249-0.192992\imo\\
$\ell=3$,WKB6&1.350732-0.193001\imo&1.350850-0.192884\imo&1.350739-0.192995\imo&1.344377-0.191715\imo&1.350569-0.192969\imo\\
\hline
$\ell=4$,WKB3&1.734680-0.192793\imo&1.734662-0.192802\imo&1.734679-0.192793\imo&1.726554-0.191509\imo&1.734471-0.192760\imo\\
$\ell=4$,WKB6&1.734831-0.192784\imo&1.734848-0.192750\imo&1.734834-0.192782\imo&1.726704-0.191500\imo&1.734622-0.192752\imo\\
\hline
$\ell=5$,WKB3&2.119141-0.192678\imo&2.119127-0.192688\imo&2.119140-0.192678\imo&2.109235-0.191395\imo&2.118886-0.192646\imo\\
$\ell=5$,WKB5&2.119224-0.192674\imo&2.119219-0.192665\imo&2.119224-0.192673\imo&2.109316-0.191391\imo&2.118969-0.192641\imo\\
\hline
\end{tabular}
\end{table*}


For a wide class of problems of astrophysical interest, the
dynamical wave equation has the form (\ref{wavelike}), where the effective
potential has the form of some potential barrier that approaches
constant values at the event horizon and spatial infinity.
The general solution of the wave equation at infinity is
\begin{equation}
\Psi = A_{in} \psi_{in} + A_{out} \psi_{out}, \quad r_{*} \rightarrow \infty.
\end{equation}
The quasi-normal modes in this approach, by the definition, are the
poles of the reflection coefficients $A_{out}/A_{in}$. Out
starting point is suggested by the WKB method \cite{WKB} where the
asymptotic solutions of the wave equation near the event horizon
and near spatial infinity are matched with Taylor expansion near
the peak of the potential, i.e. between two turning points $V(r) -
\omega^{2} = 0$. We state that {\it the low laying quasi-normal modes are
determined mainly by the behavior of the effective potential near
its peak, while the behavior of the potential far from black hole
is insignificant}. Below we shall show that our statement is true.
For this to happen, we shall consider the well-known potential for
Schwarzschild black hole $V(r)$ and also two other potentials
which lay closely to the  Schwarzschild potential near its
maximum, but has very different behavior far from a black hole.
These two potentials are chosen in the following way. We make a
plot of the function $V(r)$ of an analytic Schwarzschild
potential, then we find some number of points for this potential
$V_{r_{i}}$ near its maximum which serve us a basis for our first
potential $V_{int}$ which is an {\it interpolation} of these
points near the maximum by cubic splines. Second potential
$V_{fit}$ is a {\it fit} of the above plot near the maximum by a
ratio of polynomial functions. The possibility of making use of
the numerical interpolation of the potential near its maximum
gives us possibility of finding QN spectrum of fields near the
solutions {\it which are not known in analytical form}.

\begin{figure}\label{spot}
\caption{Potential for electromagnetic perturbations near
  Schwarzschild black hole ($\ell=2$) and the same potential interpolated
  numerically near its maximum. Even despite the behavior of the two
  potentials are very different in the full region of
  $r$, except for a small region near black hole, low-laying quasi-normal modes
  for both potentials are very close.}
\resizebox{\linewidth}{!}{\includegraphics*{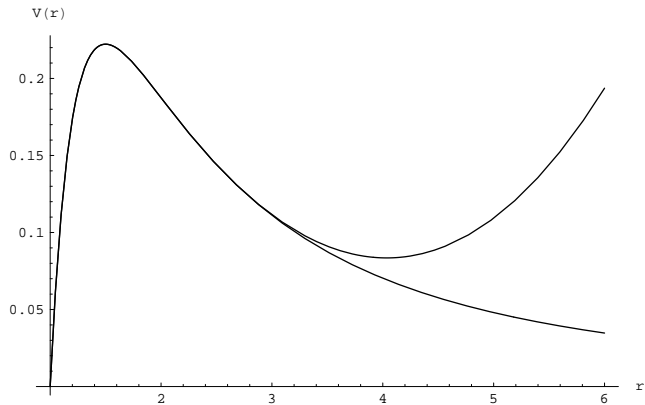}}
\end{figure}

Let us explain the interpolation technique more precisely. Suppose we know some numerical
solution for the metric. In other words, we are able to calculate
the metric coefficients in a finite number of points with desired
precision. Then, by using formulas (\ref{sp},\ref{emp}), one can
find the values of the potential at these points with the desired
precision and any number of its derivatives. It is clear that the
precision of the higher derivatives decrease very rapidly, so
that, in order to find them with a good accuracy we have to find a
lot of interpolation points very precisely. If the accuracy is not
good enough, the calculations lead to some random values. Thus we
feel that the potential has to be very smooth (and for physically
relevant problems it is) to use the WKB approximation, because WKB
method requires some number of the derivatives at the maximum of
the effective potential. The 6-th order WKB formula reads
\begin{equation}\label{WKBformula}
\frac{\imath Q_{0}}{\sqrt{2 Q_{0}''}}
- \sum_{i=2}^{i=6} \Lambda_{i} = n+\frac{1}{2},
\end{equation}
where the correction terms of the i-th WKB order $ \Lambda_{i}$
can be found in \cite{Will-Schutz} and \cite{Konoplya-prd3}, $Q =
V - \omega^2$ and $Q_{0}^{i}$ means the i-th derivative of $Q$ at
its maximum. For the WKB formula of sixth order one needs twelve
derivatives of the potentials.

Now let us see how the above programs of fit and interpolation work for the well- known
case of the Schwarzschild black hole. In cases when we know the
metric functions $N(r)$ and $B(r)$ numerically in some region near
the black hole using these functions in the formulas for
potentials (\ref{sp}-\ref{emp}) gives us numerical values for
the potential near the black hole. The accurate plot of the effective
potential (\ref{emp}) $\ell=2$ is shown on Fig.
\ref{spot} together with interpolated potential $V_{int}$ and fit
potential $V_{fit}$ where the interpolation is made on the basis
of 75 points from $r=2M$ until $r=5M$ and fit is made on the basis
of 100 points found with 1\% precision. Fit functions are chosen
as fractions of two polynomials:
\begin{equation}\nonumber
N(r) =
\frac{\displaystyle\sum_{i=0}^{i=N_N}a^{(N)}_ir^{i}}{\displaystyle1 +
  \sum_{i=1}^{i=N_N}b^{(N)}_i r^{i}}, \quad B(r) =
\frac{\displaystyle\sum_{i=0}^{i=N_B}a^{(B)}_ir^{i}}{\displaystyle1 +
  \sum_{i=1}^{i=N_B}b^{(B)}_i r^{i}},
\end{equation}
which are substituted into equations (\ref{sp}) and (\ref{emp}).
The numbers $N_N$ and $N_B$ determine the number of terms in the
polynomials and are chosen in order to provide best convergence of
the WKB series. Coefficients $a^{(N)}_i$, $b^{(N)}_i$,
$a^{(B)}_i$, $b^{(B)}_i$ are determined by the fitting procedure.
We see from the above plot, that at larger $r$ the interpolation
potential $V_{int}$ grows, while the precise function and fit
function are decreasing. In the region near black hole  $r = (2 M,
5 M)$ the difference between the potentials is negligible. The
interpolation was performed by cubic spline with the help of {\it
Mathematica\textregistered}. In order to diminish accumulating of
an error, upon calculating of the first derivative of the
interpolated potential, we had to interpolate also the obtained
values for the first derivative and all twelve consequent
derivatives up to the sixth order.

Let us look at the tables \ref{edata}-\ref{sdata}. The known
6th-order WKB quasi-normal  frequencies are compared with those
obtained by using formula (\ref{WKBformula}) for interpolation
function $V_{int}$ and for fit function $V_{fit}$. Within our
approach we have two kinds of errors: error due-to WKB
approximation, and error due-to approximation of the potential by
fit or interpolation. The error because of the WKB approximation
can be estimated simply by comparison of the QN values within
different WKB orders. The initial deviations from the accurate
potential due-to fit or interpolation induce larger errors for
higher derivatives of the potential, and thereby, show themselves
as a final error in the computed QNMs. Therefore, if one gets good
convergence of the WKB series, it means that the deviations of the
approximated potential from the accurate one are small enough, so
that induced errors in higher derivatives are small. Thus, the
criterium of goodness of a fit or interpolation is convergence of
the WKB series. An interpolation, to give a good WKB convergence,
can be improved in the following way: one should increase the
density of interpolation points, and, at the same time, increase
the accuracy with which each point is given. Unlike interpolation,
the fit approximation is more economic and works better with
increasing either number of fit points or accuracy with which each
point is given.

Using the above mentioned procedure of improving of the fit and
interpolation, we can, in principle, decrease the error due-to
potential approximation until as small value as necessary. From tables
\ref{edata}-\ref{sdata}, we see that larger number of points
for interpolation or fit gives better accuracy as compared with
data for the exact potential. While for interpolation we had to
increase precision of the potential values, for fitting it was enough
to use the fixed precision of two digits after point for simple 
illustrative Schwarzschild case.  
For instance, for $\ell =1$, $s=1$ mode
the difference between the QN frequency for the exact potential
and for the fit potential is $0.011\%$ for $Re \omega$ and
$0.016\%$ for $Im \omega$. The same order of error holds for
higher multipoles.  For lower multipoles, the error due-to fit
approximation is two orders smaller than error due-to WKB
technique. Therefore, when interpreting our new results in the
next section, for Einstein-Aether theory, we may neglect error
due-to potential approximation.

\section{Quasi-normal modes for black holes in Einstein-Aether theory}


In  the Einstein-Aether theory we know the metric functions $N(r)$
and $B(r)$ only numerically, as a result of some numerical
integration in some region near the black hole \cite{Jacobson}.
In order to integrate equation for metric coefficients with
required precision, we had to find the accurate values of
$f'(r_{h})$ with the help of shooting method described in
\cite{Jacobson}. Here we shall use the fit approximation, because, as explained
above, it works better for this case.
In this section we used for our fit $10000$ points given 
with the six digit precision after the point.  
This is 100 times more points given with accuracy four orders
better than that used in previous section for the Schwarzschild case.

The polynomial fit for the metric or the effective potential is
characterized by the number of terms in its multiplicator and
denominator for metric coefficients $N(r)$ and $B(r)$, called here
$N_{N}$ and $N_{B}$ respectively. There is some optimal number for
which the convergence of WKB series is best. Practically we
continue in the following way: in order to find optimal $N_{N}$ we
search for minimal difference between third and sixth order WKB
values, first for the case  $B(r)=1$.
When we know an optimal value  $N_{N}$,
in a similar fashion, i.e., by looking for best WKB convergence,
we are in position to find the optimal fit for $B(r)$. Quick WKB
convergence shows that higher derivatives of the metric
coefficients are calculated with the best accuracy.

\begin{table*}
\caption{Fundamental QN frequencies of the non-reduced theory for
$s=0$}\label{tresbeg}
\begin{tabular}{|l|c|c|c|c|c|}
\hline
$c_1$&$\ell=1$&$\ell=2$&$\ell=3$&$\ell=4$&$\ell=5$\\
\hline
$0.1 $&$0.589693-0.197691\imo$&$0.973933-0.195722\imo$&$1.360108-0.195176\imo$&$1.746918-0.194949\imo$&$2.134014-0.194833\imo$\\
$0.2 $&$0.593814-0.199635\imo$&$0.981227-0.197457\imo$&$1.370409-0.196886\imo$&$1.760201-0.196648\imo$&$2.150271-0.196527\imo$\\
$0.3 $&$0.598713-0.203470\imo$&$0.989398-0.201414\imo$&$1.381943-0.200829\imo$&$1.775082-0.200580\imo$&$2.168489-0.200452\imo$\\
$0.4 $&$0.604163-0.207262\imo$&$0.998783-0.205259\imo$&$1.395215-0.204663\imo$&$1.792215-0.204405\imo$&$2.189472-0.204271\imo$\\
$0.5 $&$0.610513-0.212271\imo$&$1.009919-0.210209\imo$&$1.410990-0.209603\imo$&$1.812598-0.209335\imo$&$2.214446-0.209195\imo$\\
$0.6 $&$0.618482-0.218866\imo$&$1.023707-0.216876\imo$&$1.430563-0.216280\imo$&$1.837918-0.216006\imo$&$2.245488-0.215860\imo$\\
$0.7 $&$0.628421-0.229250\imo$&$1.042296-0.226654\imo$&$1.457064-0.226066\imo$&$1.872255-0.225787\imo$&$2.287621-0.225637\imo$\\
$0.77$&$0.637216-0.240653\imo$&$1.060028-0.236792\imo$&$1.482451-0.236224\imo$&$1.905212-0.235948\imo$&$2.328101-0.235798\imo$\\
\hline
\end{tabular}
\end{table*}

\begin{table*}
\caption{Fundamental QN frequencies of the non-reduced theory for $s=1$}
\begin{tabular}{|l|c|c|c|c|c|}
\hline
$c_1$&$\ell=1$&$\ell=2$&$\ell=3$&$\ell=4$&$\ell=5$\\
\hline
$0.1 $&$0.499500-0.187126\imo$&$0.921357-0.192128\imo$&$1.322831-0.193366\imo$&$1.718010-0.193859\imo$&$2.110397-0.194105\imo$\\
$0.2 $&$0.503184-0.188782\imo$&$0.928321-0.193764\imo$&$1.332882-0.195021\imo$&$1.731094-0.195524\imo$&$2.126489-0.195775\imo$\\
$0.3 $&$0.506385-0.192059\imo$&$0.935562-0.197618\imo$&$1.343774-0.198918\imo$&$1.745483-0.199429\imo$&$2.144308-0.199683\imo$\\
$0.4 $&$0.510456-0.195179\imo$&$0.944082-0.201334\imo$&$1.356438-0.202690\imo$&$1.762147-0.203217\imo$&$2.164908-0.203477\imo$\\
$0.5 $&$0.515081-0.199198\imo$&$0.954092-0.206114\imo$&$1.371426-0.207550\imo$&$1.781923-0.208800\imo$&$2.189388-0.208370\imo$\\
$0.6 $&$0.520683-0.204318\imo$&$0.966339-0.212544\imo$&$1.389924-0.214116\imo$&$1.806415-0.214705\imo$&$2.219757-0.214992\imo$\\
$0.7 $&$0.527472-0.211949\imo$&$0.982634-0.221933\imo$&$1.414826-0.223723\imo$&$1.839523-0.224382\imo$&$2.260890-0.224700\imo$\\
$0.77$&$0.533464-0.219667\imo$&$0.997944-0.231610\imo$&$1.438525-0.233670\imo$&$1.871181-0.234419\imo$&$2.300312-0.234780\imo$\\
\hline
\end{tabular}
\end{table*}

\begin{table*}
\caption{Fundamental QN frequencies of the non-reduced theory for $s=0$, $\ell=0$}\label{tresend}
\begin{tabular}{|l|c|c|c|c|}
\hline
$c_1$& P\"{o}schl-Teller & WKB3 & WKB5  & WKB6  \\
\hline
$0.1 $&$0.230507-0.232576\imo$&$0.209816-0.233510\imo$&$0.210764-0.213968\imo$&$0.221205-0.203869\imo$\\
$0.2 $&$0.231194-0.234727\imo$&$0.209789-0.235548\imo$&$0.212313-0.215742\imo$&$0.222606-0.205766\imo$\\
$0.3 $&$0.232538-0.240341\imo$&$0.210476-0.241535\imo$&$0.212397-0.220352\imo$&$0.224395-0.208570\imo$\\
$0.4 $&$0.233729-0.245740\imo$&$0.210540-0.247090\imo$&$0.215188-0.225894\imo$&$0.222845-0.218132\imo$\\
$0.5 $&$0.235007-0.252888\imo$&$0.209718-0.254241\imo$&$0.220603-0.233407\imo$&$0.220265-0.233765\imo$\\
$0.6 $&$0.236272-0.262898\imo$&$0.207228-0.264113\imo$&$0.229154-0.243557\imo$&$0.217124-0.257051\imo$\\
$0.7 $&$0.237276-0.278203\imo$&$0.201317-0.279282\imo$&$0.238603-0.255254\imo$&$0.224289-0.271544\imo$\\
$0.77$&$0.237259-0.294753\imo$&$0.193652-0.296415\imo$&$0.240916-0.263198\imo$&$0.241130-0.262964\imo$\\
\hline
\end{tabular}
\end{table*}



From tables \ref{tresbeg}-\ref{tresend} we can see that real and
imaginary parts of $\omega$ is about $1\%$ greater for
$c_{1}=0.1$. When $c_{1}$ is growing up to $c_{1}=0.77$, both $Re
\omega$ and $Im
\omega$ are growing by about $10\%$ and $20\%$ respectively. For
$\ell \geq 1$, the WKB method gives an error of about a fraction
of a percent. Therefore our conclusion, that in the above region
of $c_{1} \leq 0.77$, the observed here effect of a few percents
is much larger than the order of the WKB error. For the particular
case of $s = \ell =0$, the WKB error is about $5\%$ (see Table
\ref{tresend}), therefore we cannot judge about the value of the
aether effect in the case $s = \ell =0$ for small $c_{1}$. Yet for
$c_{1} \sim 0.7 $, the effect a few times greater than the error,
and, qualitatively the behavior of the QN spectrum is the same:
the larger $c_{1}$ leads to a larger real oscillation frequency
and damping rate.


\section{Discussion}

The quasinormal modes of black holes should be feasible for observation
by new generation of gravitational antennas. This would make a
great impact on gravitational physics, where gravitational waves have not been not observed.

At the same time, the gravitational consequences of Local Lorentz
symmetry violation must show itself in radiative processes around
black holes. Gravitational radiation damping of binary pulsars
orbits reproduces the weak field general relativity at the lowest
post- Newtonian order \cite{Foster}.  The significant difference
between Einstein and Einstein-Aether theories can show itself in
the regime of strong field, i.e. in changing the characteristic
quasi-normal spectrum of black holes. Thereby, aether, if exists,
could be indirectly observed through observation of characteristic
spectrum of black holes. This motivated us to perform the present
study of quasi-normal modes for black holes in Einstein-Aether
theory.

In this paper we developed a new method for finding quasi-normal
modes for potentials determined only numerically in some region
near a black hole. Then, the method has been applied here to
numerical black hole solution in Einstein-Aether theory, which 
has been obtained recently in \cite{Jacobson}.

We showed that quasi-normal modes of Einstein-Aether black holes
have larger damping rate and real oscillation frequency than QNMs
of Schwarzschild black holes have. We considered here scalar and
electromagnetic test fields in the background of a spherically
symmetric black hole. The wave equation for gravitational
perturbations could be found by consideration of perturbations
of the complete system of Einstein-Aether equations. Then our approach
for finding of the quasi-normal modes can be applied as well. Yet, it
is well known, that quasinormal modes in the eikonal
approximation are the same for fields of different spin. Therefore
we expect qualitatively the same results for gravitational
perturbations as we observed for scalar and electromagnetic
perturbations. If the breaking of Lorentz symmetry is not very
small, i.e. $c_{1}$ is not very small, the deviation of the QN
modes from their Schwarzschild values might be observed in the
forthcoming experiments with gravitational antennas.


\begin{acknowledgments}
This work was supported by \emph{Funda\c{c}\~{a}o de Amparo
\`{a} Pesquisa do Estado de S\~{a}o Paulo (FAPESP)}, Brazil.
We would like to thank  C. Eling for sharing with us his numerical
code for integration of the Einstein-Aether equations.
\end{acknowledgments}


\begin{thebibliography}{80}
\bibitem{LV}
  F.~Ahmadi, S.~Jalalzadeh and H.~R.~Sepangi,
  arXiv:gr-qc/0605038;
  Q.~G.~Bailey and V.~A.~Kostelecky,
  arXiv:gr-qc/0603030;
G.L. Alberghi, R. Casadio, A. Tronconi, hep-ph/0310052;
  V.~A.~Kostelecky and R.~Potting,
  Gen.\ Rel.\ Grav.\  {\bf 37}, 1675 (2005)
  [Int.\ J.\ Mod.\ Phys.\ D {\bf 14}, 2341 (2005)]
  [arXiv:gr-qc/0510124];
  B.~Altschul,
  Phys.\ Rev.\ D {\bf 72}, 085003 (2005)
  [arXiv:hep-th/0507258];
  T.~G.~Rizzo,
  JHEP {\bf 0509}, 036 (2005)
  [arXiv:hep-ph/0506056];
  T.~Jacobson, S.~Liberati and D.~Mattingly,
  Annals Phys.\  {\bf 321}, 150 (2006)
  [arXiv:astro-ph/0505267];
  S.~Groot Nibbelink and M.~Pospelov,
  Phys.\ Rev.\ Lett.\  {\bf 94}, 081601 (2005)
  [arXiv:hep-ph/0404271].
\bibitem{AEreview}
  C.~Eling, T.~Jacobson and D.~Mattingly,
  arXiv:gr-qc/0410001.
\bibitem{AEobserve}
  C.~Heinicke, P.~Baekler and F.~W.~Hehl,
  Phys.\ Rev.\ D {\bf 72}, 025012 (2005)
  [arXiv:gr-qc/0504005];
  M.~L.~Graesser, A.~Jenkins and M.~B.~Wise,
  Phys.\ Lett.\ B {\bf 613}, 5 (2005)
  [arXiv:hep-th/0501223];
  R.~Bluhm and V.~A.~Kostelecky,
  Phys.\ Rev.\ D {\bf 71}, 065008 (2005)
  [arXiv:hep-th/0412320];
  T.~Jacobson, S.~Liberati and D.~Mattingly,
  Lect.\ Notes Phys.\  {\bf 669}, 101 (2005)
  [arXiv:hep-ph/0407370].
  B.~Z.~Foster,
  Phys.\ Rev.\ D {\bf 73}, 024005 (2006)
  [arXiv:gr-qc/0509121];
  B.~Z.~Foster and T.~Jacobson,
  Phys.\ Rev.\ D {\bf 73}, 064015 (2006)
  [arXiv:gr-qc/0509083];
  C.~M.~L.~de Aragao, M.~Consoli and A.~Grillo,
  arXiv:gr-qc/0507048;
  T.~Jacobson and D.~Mattingly,
  Phys.\ Rev.\ D {\bf 70}, 024003 (2004)
  [arXiv:gr-qc/0402005];
\bibitem{CFT}
G. T. Horowitz and V. E. Hubeny,  Phys. Rev. D \textbf{62}, 024027 (2000);
  D.~Birmingham, I.~Sachs and S.~N.~Solodukhin,
  Phys.\ Rev.\ Lett.\  {\bf 88}, 151301 (2002)
  [arXiv:hep-th/0112055];
  A.~O.~Starinets,
  Phys.\ Rev.\ D {\bf 66}, 124013 (2002)
  [arXiv:hep-th/0207133];
  V.~Cardoso, R.~Konoplya and J.~P.~S.~Lemos,
  Phys.\ Rev.\ D {\bf 68}, 044024 (2003)
  [arXiv:gr-qc/0305037];
  R.~A.~Konoplya,
  Phys.\ Rev.\ D {\bf 66}, 044009 (2002)
  [arXiv:hep-th/0205142];
  V.~Cardoso and J.~P.~S.~Lemos,
  Phys.\ Rev.\ D {\bf 63}, 124015 (2001)
  [arXiv:gr-qc/0101052].
\bibitem{QNMs}
  A.~Lopez-Ortega,
  arXiv:gr-qc/0605034.
  A.~Lopez-Ortega,
  arXiv:gr-qc/0605027.
  P.~Kanti and R.~A.~Konoplya,
  Phys.\ Rev.\ D {\bf 73}, 044002 (2006)
  [arXiv:hep-th/0512257];
  S.~Musiri, S.~Ness and G.~Siopsis,
  Phys.\ Rev.\ D {\bf 73}, 064001 (2006)
  [arXiv:hep-th/0511113];
  E.~Abdalla, R.~A.~Konoplya and C.~Molina,
  Phys.\ Rev.\ D {\bf 72}, 084006 (2005)
  [arXiv:hep-th/0507100];
  A.~Ghosh, S.~Shankaranarayanan and S.~Das,
  Class.\ Quant.\ Grav.\  {\bf 23}, 1851 (2006)
  [arXiv:hep-th/0510186];
  P.~P.~Fiziev,
  Class.\ Quant.\ Grav.\  {\bf 23}, 2447 (2006)
  [arXiv:gr-qc/0509123];
  R.~Konoplya,
  Phys.\ Rev.\ D {\bf 71}, 024038 (2005)
  [arXiv:hep-th/0410057];
  S.~Das and S.~Shankaranarayanan,
  Class.\ Quant.\ Grav.\  {\bf 22}, L7 (2005)
  [arXiv:hep-th/0410209];
  C.~G.~Shao, B.~Wang, E.~Abdalla and R.~K.~Su,
  Phys.\ Rev.\ D {\bf 71}, 044003 (2005)
  [arXiv:gr-qc/0410025];
  R.~A.~Konoplya and E.~Abdalla,
  Phys.\ Rev.\ D {\bf 71}, 084015 (2005)
  [arXiv:hep-th/0503029].
  A.~J.~M.~Medved, D.~Martin and M.~Visser,
  Class.\ Quant.\ Grav.\  {\bf 21}, 2393 (2004)
  [arXiv:gr-qc/0310097];
  T.~Padmanabhan,
  Class.\ Quant.\ Grav.\  {\bf 21}, L1 (2004)
  [arXiv:gr-qc/0310027];
  R.~A.~Konoplya,
  Phys.\ Rev.\ D {\bf 68}, 124017 (2003)
  [arXiv:hep-th/0309030];
  Phys.\ Rev.\ D {\bf 73}, 024009 (2006)
  [arXiv:gr-qc/0509026];
  M.~R.~Setare Class.\ Quant.\ Grav. {\bf 21} 1453 (2004) [hep-th/0311221];  Phys.\ Rev.\ D{\bf 69} 044016 (2004) [hep-th/0312061];
  R.~A.~Konoplya and A.~Zhidenko,
  arXiv:gr-qc/0605013;
  S.~Fernando,
  Gen.\ Rel.\ Grav.\  {\bf 36}, 71 (2004)
  [arXiv:hep-th/0306214];
\bibitem{Jacobson}
  C.~Eling and T.~Jacobson,
  arXiv:gr-qc/0604088;


  C.~Eling and T.~Jacobson,
  arXiv:gr-qc/0603058.
\bibitem{Zhidenko:2005mv}
  A.~Zhidenko,
  %
  Class.\ Quant.\ Grav.\  {\bf 23}, 3155 (2006)
  [arXiv:gr-qc/0510039].
\bibitem{WKB}B.F.Schutz and C.M.Will  {\it Astrophys.J.Lett}  {\bf{291}} L33
(1985);
\bibitem{Will-Schutz} S.Iyer and C.M.Will,  {\it Phys.Rev.}  {\bf{D35}} 3621 (1987)
\bibitem{Konoplya-prd3} R. A. Konoplya, Phys.Rev D {\bf 68}, 024018
(2003);
  R.~A.~Konoplya,
  J.\ Phys.\ Stud.\  {\bf 8}, 93 (2004);
\bibitem{Foster}
  B.~Z.~Foster,
  arXiv:gr-qc/0602004;
\end{thebibliography}
\end{document}